\documentclass[a4paper,11pt]{article}
\usepackage{pos}
\usepackage{natbib}
\usepackage{subfig}
\usepackage{verbatim}
\graphicspath{ {figures/} }

\title{Scattering of SIMPlectic Dark Pions}

\author*[a]{Yannick Dengler}
\author[a]{Axel Maas}
\author[b]{Fabian Zierler}

\affiliation[a]{University of Graz, Universitätsplatz 5, 8010 Graz, Austria}
\affiliation[b]{Swansea University, Singleton Park, SA2 8PP Swansea, Wales, United Kingdom}

\emailAdd{yannick.dengler@uni-graz.at}
\emailAdd{axel.maas@uni-graz.at}
\emailAdd{fabian.zierler@swansea.ac.uk}

\abstract{
We study the scattering of two identical pNGBs (\textbf{p}seudo-\textbf{N}ambu-\textbf{G}oldstone \textbf{B}osons) in $Sp(4)$ gauge theory with two mass-degenerate Dirac fermions in the pseudo-real fundamental representation. This theory serves as a realization of a SIMP (\textbf{S}trongly \textbf{I}nteracting \textbf{M}assive \textbf{P}articles) Dark Matter model. SIMPs  are an exciting dark matter candidate as they make use of a new relic density mechanism and provide potential solutions to the so-called small-scale structure problems. These theories are realized by a confining dark sector which includes non-perturbative signatures. While most of the research focuses on ChPT (\textbf{Ch}iral \textbf{P}erturbation \textbf{T}heory), first-principle verification of these models is indispensable. In these proceedings we give an update on scattering properties in the most common channel and give an outlook on the projects that lie ahead.}

\FullConference{
The 41th International Symposium on Lattice Field Theory (Lattice 2024)\\
July 28st - August 3rd, 2024\\
Liverpool, UK
}

\begin{document}
\maketitle

\section{Introduction}

Beyond the standard model sectors described by symplectic gauge groups with fundamental fermions have interesting phenomenological properties for composite Higgs or dark matter models \cite{Bennett:2023wjw,Kulkarni:2022bvh}. 
In contrast to SU(N), Sp(2N) has a pseudo-real fundamental fermion representation. Left- and right-handed spinor components are related by the global flavour symmetry which is enlarged compared to SU(N) theories \cite{Kogut:2000ek}.

We use this setup to describe a dark sector in which the dark matter candidates inherit a self-interaction. 
This possibility gained attention recently as it provides a possible solution to the so-called small-scale structure problems. Moreover, this model is a minimal realization of the so-called SIMP paradigm in which the dark matter relic abundance is realized making use of a number-lowering process within the dark sector \cite{Hochberg:2014kqa, Hochberg:2014dra}. Calculations of i.e. the relic density in SIMP theories have been done using ChPT. 
The lattice can provide first-principle input to these calculations in the form of LECs (\textbf{L}ow-\textbf{E}nergy \textbf{C}onstants). This is necessary as the underlying UV (\textbf{U}ltra \textbf{V}iolet) theory is inherently non-perturbative.
Charging two fundamental fermions under the fundamental representation of Sp(4), results in a U(4)$_F$ flavour symmetry. The breaking of the axial anomaly leaves an SU(4)$_F$ flavour symmetry. Chiral symmetry breaking and explicit fermion masses break this symmetry further down to Sp(4)$_F$.

This breaking gives rise to five pNGBs\footnote{The number of broken generators or the dimension of the coset SU(4)/Sp(4)} which will be the lightest particle in the theory and our dark matter candidate. In similarity to QCD, we will refer to them as \textit{dark pions}. 

These five pions can interact via a Wess-Zumino-Witten 5-point-vertex \cite{Wess:1971yu,Witten:1983tw} in ChPT and therefore accommodate the number lowering process relevant for the SIMP paradigm.

In a first step, the mass spectrum in the theory was investigated \cite{Bennett:2019jzz}, but for the investigation of the theory as a dark matter candidate, scattering information is essential. In Ref.~\cite{Dengler:2024maq}, of which we will show result in these proceedings, we investigated the scattering phase shift in the maximal scattering channel. As a first step, this work already gave the confirmation that the theory passes experimental bounds on i.e. the pion mass. In ongoing efforts, we perform lattice calculations closer to the chiral limit, which has the advantage that lattice observables can be translated to LECs more reliably. Further, this increases the ratio of the vector particle mass to the pion mass, eventually making former resonant in the 10-dimensional scattering channel, which will have important phenomenological consequences \cite{Bernreuther:2023kcg}. 
A milestone in Sp(4) dark matter is to calculate the amplitude of the 5-point vertex which also lives in the 10-dimensional channel, checking whether an effective description is viable and calculating the resulting relic density which will be the ultimate test for the model as a dark matter candidate. 

For further details, we refer to our paper \cite{Dengler:2024maq}.  

\section{Lattice setup}

We are interested in first-principle spectroscopy and scattering information. To this end, we use lattice field theory and the Lüscher quantization condition \cite{Luscher:1985dn,Luscher:1986pf}. In this chapter, we state the techniques employed and present some details and improvements that are not found in \cite{Dengler:2024maq}.

For our calculations we use the HiRep code \cite{DelDebbio:2008zf, HiRepSUN} which has been extended to simulations of symplectic gauge groups \cite{HiRepSpN}. We generate ensembles using the unimproved Wilson action with dynamical fermions. We explore a parameter space in the inverse gauge coupling $\beta$ = 6.9, 7.05 and 7.2 and corresponding bare fermion masses which result in values of $\frac{m_\pi}{m_\rho}$ between 0.70 and 0.88. 

We calculate correlation functions of a single pion and a two-pion scattering operator in the most common channel that are given by

\begin{align}    
\label{Wick_Operators}
\begin{split}
    \mathcal{O}_{\pi}(x) &= \bar{u}(x)\gamma_5 d(x) \\
    \mathcal{O}_{\pi\pi}(x,y) &= \mathcal{O}_{\pi}(x)\mathcal{O}_{\pi}(y)= \bar{u}(x) \gamma_5 d(x) \bar{u}(y) \gamma_5 d(x).
\end{split}
\end{align}

Performing the Wick contractions, the two-pion correlation function receives contributions from two diagrams $C_{\pi\pi} = 2D-2C$, where "D" and "C" stand for \textit{disconnected} and \textit{connected} respectively. 

We use $Z_2 \times Z_2$ stochastic noise sources with spin-dilution \cite{Foley:2005ac} for the inversion of the Dirac operator. In order to remove constant contributions to the correlation function from \textit{around-the-world} effects, we perform a numerical derivative \cite{Umeda:2007hy} before fitting the data using the corrfitter package \cite{peter_lepage_2021_5733391}. It uses a Bayesian approach to fit a tower of energy levels. In the final analysis, we only use the first energy level. 

ChPT expands order by order in $\frac{m_\pi}{f_\pi}$. In order to compare our results with ChPT, we calculate the unrenormalized pion decay constant $f_\pi^0$ via
\begin{align}\label{eq:pion_decay_constant}
    \lim_{t\to\infty} C_{\mathcal{O}_{\gamma_0 \gamma_5}}(t) = \frac{\vert \langle 0 \vert \mathcal O_{\gamma_0 \gamma_5} \vert {\rm PS} \rangle \vert^2}{2m_\pi} \left(  e^{-m_\pi t} + e^{-m_\pi (T-t)}  \right) = \frac{\left( f^0_\pi \right)^2 m_\pi}{2} \left(  e^{-m_\pi t} + e^{-m_\pi (T-t)}  \right).
\end{align}
Before determining the renormalized decay constant via $f_\pi = Z_A f_\pi^0$ where the renormalization constant $Z_A$ is estimated using leading order lattice perturbation theory \cite{Martinelli:1982mw}.

\subsection{Lüscher analysis}

When two particles that posses an interaction among them are put in a finite volume, the energy levels they can take are shifted from the non-interacting ones. This shift can be correlated with the scattering behaviour in infinite volume by the so-called Lüscher quantization condition \cite{Luscher:1985dn,Luscher:1986pf}. In the most general form, the Lüscher quantization condition can be written as
\begin{equation}
    \label{eq:Luscher}
    \tan \left[ \delta(E_{\rm com}) \right] = f(E, \Vec{P}, L) |_{E=E(L)}
\end{equation}
where $\delta$ is the scattering phase shift containing all relevant scattering information as a function of the center-of-mass energy and f is a function that depends on $E$, the energy measured on the lattice, $L$ the spacial extent $P$, the total momentum. The vertical bar indicates that this formula is only valid if $E$ is an energy level in the finite volume. The extension to scattering of up to three particles \cite{Hansen:2014eka} in general frames \cite{Rummukainen:1995vs} has been worked out and will become relevant for the next steps of the investigation of Sp(4) gauge theory as a description of a dark sector. 

\section{Scattering length}

The threshold value of the phase shift is given by the scattering length. We can extract the value of the scattering phase shift by fitting it to ERE (\textbf{E}ffective \textbf{R}ange \textbf{E}xpansion) \cite{Kondo:2022lgg}. Moreover, the result can be used to fix LECs that arise in an effective description of the theory with ChPT\footnote{There was a discrepancy between the chiral perturbation theory prediction and the line drawn in Fig.~3 in \cite{Dengler:2024maq} of a factor $\pi/2$. We thank Daniil Krichevskiy for bringing this to our attention.}.

The leading order prediction for the scattering length from chiral perturbation theory for theories with two fermions charged in a pseudo-real representation of the gauge group reads \cite{Bijnens:2011fm}
\begin{align}
    \label{ChiPT}
    \begin{split}
        a_0^B &= -\frac{1}{32\pi}\left(\frac{m_\pi}{f_\pi}\right)^2 \\
        M_0^{th} &= 32\pi a_0^B=-\left(\frac{m_\pi}{f_\pi}\right)^2\\
        \sigma_0^{\text{th}}&=
        \lim_{p\to 0}\int d\Omega\frac{d\sigma}{d\Omega}=\frac{1}{64\pi}\frac{|M_0^{th}|^2}{m_\pi^2}=\frac{1}{64\pi}\left(\frac{m_\pi}{f_\pi^2}\right)^2.
    \end{split}
\end{align}
This was also used in \cite{Arthur:2014zda} with a different convention of the scattering length. The superscript $^B$ indicates the definition of the scattering length used in \cite{Bijnens:2011fm}, which differs from our definition. The third line is the threshold value for the s-wave cross-section which can be compared to the lattice data by using the ERE. In the ERE, we expand the phase shift as follows \cite{Kondo:2022lgg}
\begin{align}
    \label{eq:ChiPT}
    \begin{split}
        p \cot(\delta_0) &= -\frac{1}{a_0} +\frac{p^2 r_0}{2} + \mathcal{O}(p^4) \\
        \sigma_0&=\frac{4\pi a_0^2}{\left|1-\frac{a_0 r_0}{2}+i p a_0 \right|^2} \overset{p\to 0}{=\!=}4 \pi a_0^2 \\
        a_0 m_\pi &= \frac{1}{16\pi}\left(\frac{m_\pi}{f_\pi}\right)^2.
    \end{split}
\end{align}
The last line shows the ChPT prediction for the scattering length that we extract from the Lüscher analysis.

\section{Results}

\begin{figure}
\begin{center}
    \includegraphics[width = 0.90\textwidth]{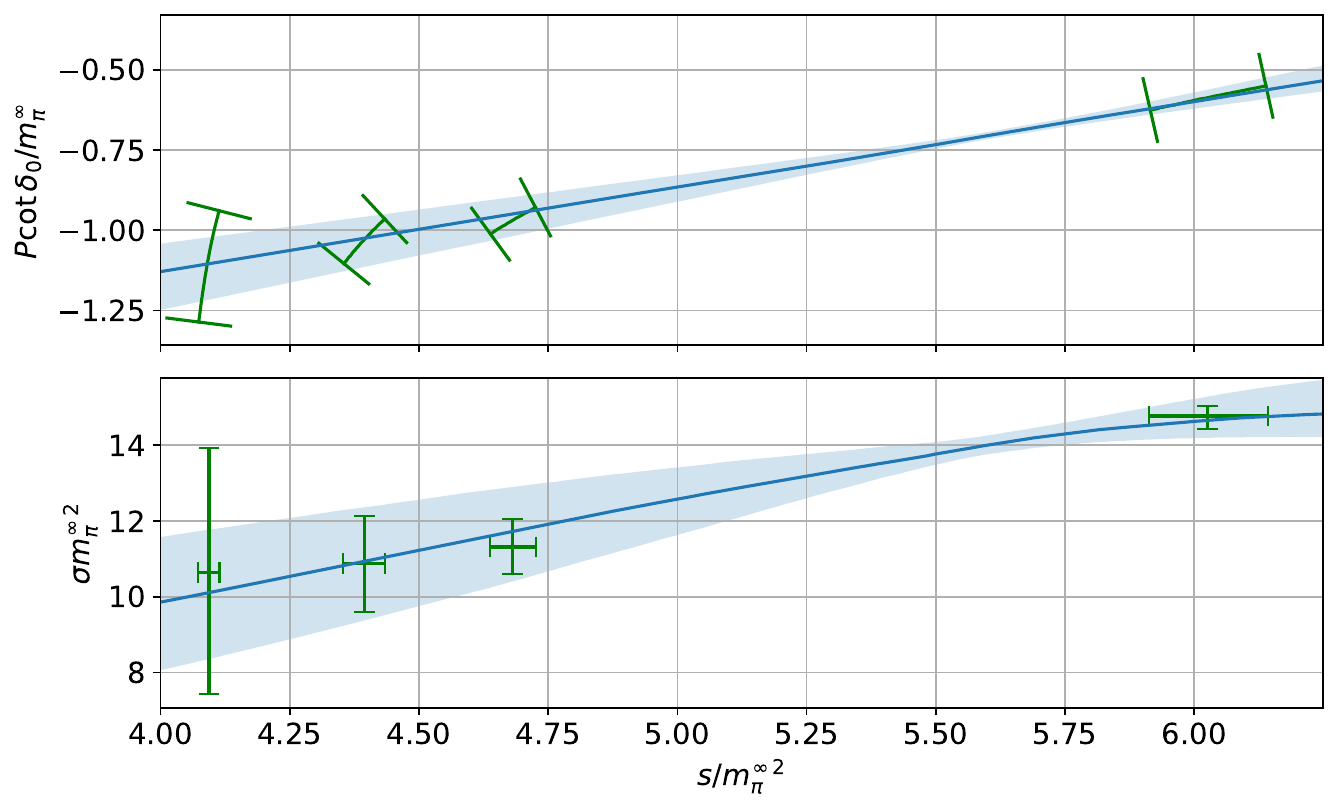}
    \caption{Top panel: Result of \eqref{eq:Luscher} plotted against the center-of-mass energy squared for one representative ensembles ($\beta$ = 7.2 and $a m_0$ = -0.78). The blue line and band show the result of fitting the data with effective range expansion. The curved error bars indicate, that the two axes are not independent and only values on the curved lines are allowed. Bottom panel: Result of using the data points above in \eqref{eq:ChiPT} to obtain the s-wave cross-section.}
    \label{plot:phase_shift}
\end{center}
\end{figure}
\begin{figure}
\begin{center}
    \includegraphics[width = 0.90\textwidth]{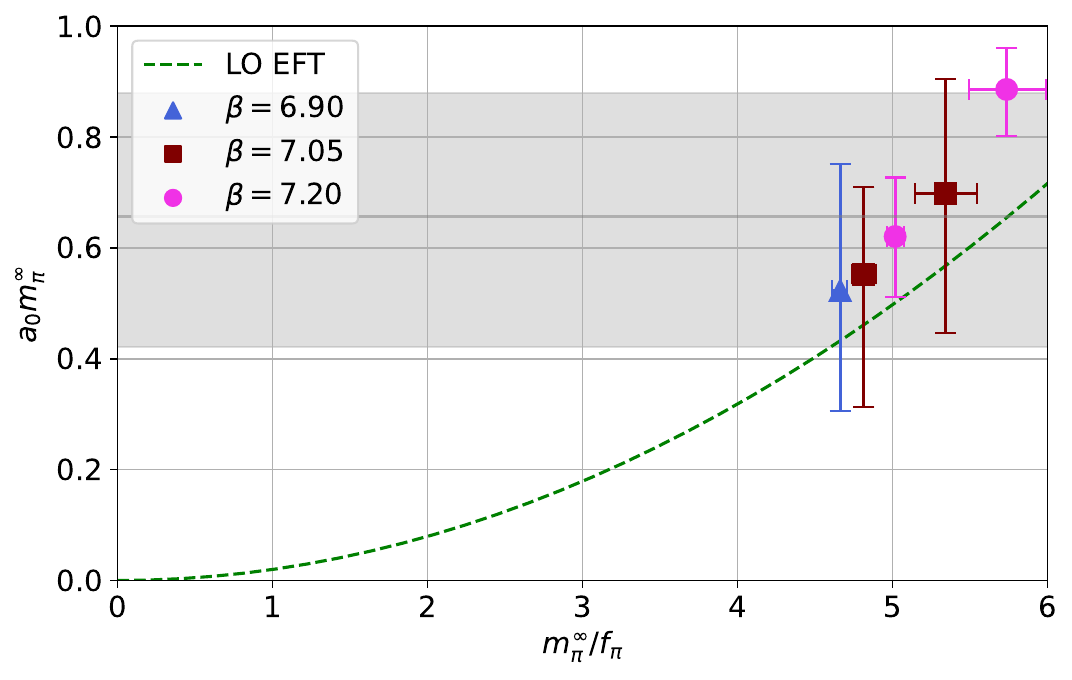}
    \caption{The scattering length obtained from effective range expansion against the ratio of the mass and decay constant of the pion (The expansion parameter in ChPT). Different colours and symbols correspond to different values of the inverse coupling $\beta$. We observe a consistent positive  scattering length across all ensembles. The horizontal grey line and band indicate a central value and error for the scattering length estimated using all of our ensembles. The green dashed line shows the expected result from leading order ChPT \cite{Bijnens:2011fm} (Note the different sign convention).}
    \label{plot:a0mpi}
\end{center}
\end{figure}

In these proceedings, we focus on the results of the scattering length. Fig.~\ref{plot:a0mpi} shows a corrected version of Fig.~3 from \cite{Dengler:2024maq} with the correct ChPT prediction on the scattering length from \cite{Bijnens:2011fm}. 
An in-depth discussion of the results can be found in \cite{Dengler:2024maq}. 

The data points in Fig.~\ref{plot:phase_shift} are the result of using the Lüscher formula for the energy levels. The curved error bars indicate that the two axes are not independent of each other, and only values on the lines are allowed. In contrast to \cite{Dengler:2024maq}, here we only consider ensembles with $N_L>8$, $m_\pi / m_\pi^\infty > 1.3$ and $aE_{\pi\pi} < 0.95$ in order to reduce finite volume and discretization artefacts. We observe consistent negative values for the phase shift at low energies which correspond to a positive scattering length in our convention and a repulsive interaction. The blue line and band show the result of fitting the data points to an effective range expansion \eqref{eq:ChiPT}. For the estimation of the error we perform a bootstrap resampling of the energy levels according to the error determined from the corrfitter package \cite{peter_lepage_2021_5733391}. We run the analysis with each sample and obtain an estimate on the error from the resulting sample. The error band is drawn, by also sampling the value of the y-axes at several values of $s/m_\pi^{\infty 2}$. The threshold value ($s\to4m_\pi^2$) can be translated to the scattering length. The bottom panel shows the s-wave cross-section obtained with \eqref{eq:ChiPT}. 

Fig.~\ref{plot:a0mpi} shows the result of the scattering length for different values of $m_\pi / f_\pi$. This plot is an updated version of Fig.~2 from \cite{Dengler:2024maq} where the ChPT prediction is now indicated by the dashed green line correctly. The data points are mostly compatible with the ChPT prediction at the $1\sigma$ level. This is promising for the use of ChPT, as one cannot naively expect the prediction of the leading order to work so well at values of $m_\pi / f_\pi\approx 4.5-6$. These values for the scattering length can further be used to describe the low-energy scattering behaviour of dark pions. In Ref.~\cite{Dengler:2024maq} this has been done to calculate the velocity-weighted cross-section which has been estimated from dark matter halos \cite{Kaplinghat:2015aga}. This showcases a direct comparison of lattice results with astrophysical data. While the lattice data does not show a velocity dependence at non-relativistic velocities, the data could still be used to constrain the mass of the dark matter candidate. We found that $m_{DM}\gtrsim 100 \,\text{MeV}$, which fits the prediction for SIMP dark matter.

\section{Outlook}

In this section I would like to summarize the problems that will arise in the future in the treatment of SIMP dark matter on the lattice. Of great importance is the interplay between effective theories and first-principle calculations. Most SIMP studies, e.g. on relic density, work with effective theories as it is much easier to describe the important degrees of freedom. However, the model is motivated by a UV-complete theory and therefore always requires first-principle verification. Correlating our results with low-energy constants in ChPT is a necessary endeavour that can already be started with the results presented here. However, to gain a thorough understanding of SIMP dark matter and in order to be able to falsify it, one needs to address the number-lowering process that motivates the SIMP paradigm in the first place. In our work, this process is realized by a $3 \to 2$ of pNGB in the second-largest scattering channel (10-dimensional). A natural first step in this channel is to measure the resonance parameters of the vector particle. This is an ongoing effort that involves new challenges, including the greater computing power required. Fortunately, all measurements of $\pi \pi \to \rho$ scattering can be repurposed to study the $3 \to 2$ process, but it remains to be seen whether it is feasible to extract enough scattering information to constrain the ChPT prediction to the $3 \to 2$ process. 

\acknowledgments

We thank Daniil Krichevskiy, Helena Kolešová and Suchita Kulkarni for the helpful discussions. We thank the authors of \cite{Bennett:2019jzz} for sharing their gauge configurations for the purposes of this paper. YD and FZ have been supported the Austrian Science Fund research teams grant STRONG-DM (FG1). FZ has been supported by the STFC Grant No. ST/X000648/1. The computations have been performed on the Vienna Scientific Cluster (VSC4).  

{\bf Research Data Access Statement}---The data generated and the analysis code for the full paper \cite{Dengler:2024maq} can be downloaded from  Ref.~\cite{data_release, code_release}. 

{\bf Open Access Statement}---For the purpose of open access, the authors have applied a Creative Commons Attribution (CC BY) license to any Author Accepted Manuscript version arising.

\bibliography{bibliography}

\end{document}